\documentclass[aip,pof,superscriptaddress,showpacs,reprint,english]{revtex4-1}
\usepackage[T1]{fontenc}
\usepackage[utf8]{luainputenc}
\usepackage{textcomp}
\usepackage{amsmath}
\usepackage{amssymb}
\usepackage{graphicx}
\usepackage{color}
\usepackage{ulem}
\makeatletter
\usepackage{dcolumn}
\usepackage{bm}
\usepackage{float}
\usepackage{mathtools}
\usepackage{lipsum}

\@ifundefined{textcolor}{}
{%
 \definecolor{BLACK}{gray}{0}
 \definecolor{WHITE}{gray}{1}
 \definecolor{RED}{rgb}{1,0,0}
 \definecolor{GREEN}{rgb}{0,0.7,0}
 \definecolor{BLUE}{rgb}{0,0,1}
 \definecolor{CYAN}{cmyk}{1,0,0,0}
 \definecolor{MAGENTA}{cmyk}{0,1,0,0}
 \definecolor{YELLOW}{cmyk}{0,0,1,0}
 }
 
\newcommand{\W}{8cm}
\newcommand{\WL}{9cm}

\begin{document}
\title{Deformation of a flexible fiber in a viscous flow  past an obstacle}
\author{H. M. L\'opez}
\email{lopezhmatias@gmail.com}
\affiliation{Grupo de Medios
Porosos, Facultad de Ingenier\'{\i}a, Paseo Colon 850, 1063, Buenos
Aires (Argentina), CONICET (Argentina).}
\author{J-P. Hulin}   
\email{hulin@fast.u-psud.fr}
\affiliation{Univ Paris-Sud, CNRS, F-91405.
  Lab FAST, B\^at 502, Campus Univ, Orsay, F-91405 (France).}
\author{H. Auradou}   
\email{Harold.Auradou@u-psud.fr}
\affiliation{Univ Paris-Sud, CNRS, F-91405.
  Lab FAST, B\^at 502, Campus Univ, Orsay, F-91405 (France).}
\author{M. V. D'Angelo}
\email{veronica.dangelo@gmail.com}
\affiliation{Grupo de Medios
Porosos, Facultad de Ingenier\'{\i}a, Paseo Colon 850, 1063, Buenos
Aires (Argentina), CONICET (Argentina).}

\begin{abstract}
We study the deformation  and transport of  elastic fibers  in a viscous Hele-Shaw  flow with curved streamlines.  The variations of the global velocity and orientation of the fiber
follow closely those of the local flow velocity. The ratios of the  curvatures of the fibers
by the corresponding curvatures of the streamlines reflect a balance between elastic and viscous forces: this ratio 
is shown experimentally to be determined by a dimensionless {\it Sperm number} $Sp$ combining the characteristic parameters 
of the flow (transverse velocity gradient, viscosity, fiber diameter/cell gap ratio) and those of the fiber (diameter, effective length, Young's modulus). For short fibers, the effective length is that of the fiber; for long ones, it is equal to the transverse characteristic
length of the flow.  For  $S_p \lesssim 250$, the ratio of the curvatures   increases  linearly with  $Sp$;  For $S_p \gtrsim 250$, the fiber reaches the same curvature as the streamlines.
\end{abstract}

\maketitle
\section{Introduction}\label{Introduction}
The transport of flexible biological or man made fibers by a flow and their deformation is of interest in view of  their potential 
applications in many different industrial fields. A first example is the paper industry  in which the orientation and spatial distributions of the fibers in  the flowing pulp must be controlled~\cite{Yasuda2004,Tornberg2006}. Fibers are also widely used  by  petroleum engineers~\cite{Howard1995} to enhance the proppant transport capabilities of fracturing fluids, or to prevent the backflow of proppant. Other domains of application  are civil engineering (special cements, structural reinforcement), textile
engineering,  bio engineering and medicine. 

In flows at  low Reynolds numbers, the deformation of  fibers reflects a balance between viscous and elastic stresses and plays often an important part: for instance, these effects are critical for understanding the dynamics of flexible biological filaments~\cite{Purcell1976,Lagomarsino2003,Lauga2009}, the filtration and cross stream migration of macromolecules in small channels \cite{Lagomarsino2004,Chelakkot2010,Reddig2011,Slowicka2012} and the positioning of bio-fibers in porous media~\cite{Rusconi2010}.
In all these applications, the prediction of the motion of the fibers is a difficult fundamental problem, particularly  in complex geometries  and  inhomogeneous flows where constrictions and obstacles are present: these create complex interactions between the flexible objects, the flowing fluids  and the solid walls. 

Our interest in these problems is  focused on  the injectability and transport of long flexible fibers through fractures in  reservoir 
rocks.  
In a previous work~\cite{Dangelo2009}, we studied   fiber transport by a flow in single narrow model rough fractures and of its dependence on  the fluid velocity, the flexibility of the fiber and the configuration of the aperture field. 

The present work is devoted to 
a quantitative study of the influence of the parameters in the configuration  where the fiber passes in the vicinity of contact areas or close to regions of low aperture.  This  situation is modeled experimentally by means of a Hele-Shaw cell of constant aperture and of width varying with the distance: this allows one to determine  the  variations  along the flow of the orientation and deformation of the flow lines and of a fiber placed in the flow. More specifically, we seek to relate the variations of the curvature of the fiber to those of the streamlines,  to the velocity gradients and to the geometrical and mechanical properties of the fiber.

Most past studies  considered the simpler case of flexible fibers in a non confined parallel shear flow~\cite{Forgacs1959,Stockie1998,Wang2006,Lindstrom2007} of shear rate $(\partial v/\partial n)$ ($v$ is the modulus of the local flow 
velocity and the derivative is taken is the direction perpendicular to the flow lines). In such a configuration, the fiber may rotate about its axis and  buckles above a  threshold value of the Sperm number  which characterizes the relative magnitudes of the viscous and elastic forces. and is equal to:
\begin{equation} \label{eq:sperm}
 \frac{\mu (\partial v/\partial n) \xi^4}{ E I};
\end{equation}
$\mu$ is the viscosity, $(\partial v/\partial n)$ is the shear rate along the fiber, $E$  its Young's modulus, $I$ its area moment of inertia ($I=\pi D^4/64$ for a circular cross section of diameter $D$) and $\xi$ a characteristic dimension (for instance the length $L_f$ of the fiber). Other studies dealt with the cross stream migration of a flexible fiber in a parallel Poiseuille flow which 
is also strongly influenced by buckling~\cite{Chelakkot2010,Reddig2011,Slowicka2012}. Finally, buckling is also observed above a critical value  $Sp \sim 120$~\cite{Wandersman2010}) of $Sp$ for fibers approaching a	 stagnation point~\cite{Young2007,Guglielmini2012,Kantsler2012}.
Numerically, slender-body theory has  been used to model the behavior of an elastic fiber (with one clamped end) in the vicinity of a corner in a curved channel~\cite{Autrusson2011}. In this latter work, the variation with time and the final shape of the filament were analyzed in different corner geometries as a function of $Sp$. 
Recently, Wexler et al.~\cite{Wexler2013}  considered a fiber with one clamped end and initially transverse to a flow confined between two plates (like in the present case). In all cases, bending results from the combined effects of the hydrodynamic and viscous forces on the fibers: yet, in Refs.~\onlinecite{Autrusson2011,Wexler2013}, their resultant is balanced by a force on the attachment point while, in the present case of a free fiber, this resultant must be zero, leading to very different shapes of the fibers.
 
In the following, we characterize first the flow field in the Hele-Shaw  cell of varying width by means of  dye tracer measurements and of $3D$ numerical simulations: we study in particular the variations of the velocity along 
the streamlines. Then, the local velocity and mean orientation of the  fibers are compared to those of the fluid velocity at the same location.
The curvature $\kappa_f$ of the fibers is characterized by its ratio $\kappa_f/\kappa_s$ to that of the streamlines: the variations of the extremal values of this ratio  on different streamlines are studied as a function of  the diameter $D$ and length $L_f$ of the fiber, of its initial transverse location and of the flow velocity. 
The definition of the Sperm number $Sp$ of the problem allowing one to take into account the influence of these different parameters is
 then discussed and the experimental dependence of $\kappa_f/\kappa_s$ on $S_p$ is analyzed.
\section{Experimental Procedure}\label{Experimental_methods}
\begin{figure}[htbp]
\includegraphics[width=\W]{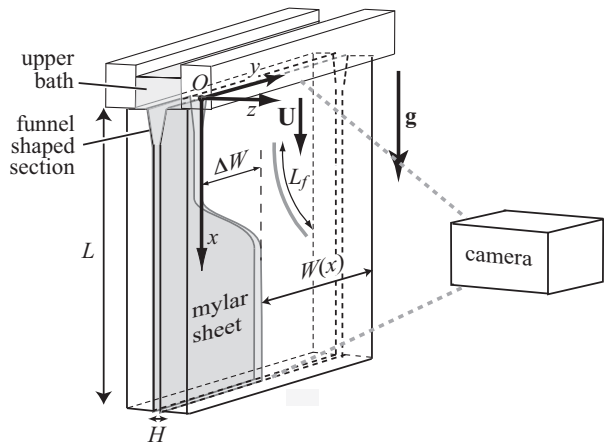}
\caption{Experimental setup.}
\label{fig:exp} 
\end{figure}

The experimental setup consists of a Hele Shaw cell placed vertically. 
Its height and gap are respectively $L=280$ and $H = 2.2 \pm 0.1\, \mathrm{mm}$ (Fig.~\ref{fig:exp}). The upper part of the cell has a Y-shaped section;  the top  is attached to a fluid bath with a slit at the bottom allowing for the flow of the fluid and the injection of the fibers. The fiber is injected in  a funnel-shaped device which permits to select conveniently the location at which the fiber is injected. A rounded off step is then cut out of a transparent mylar sheet of thickness equal to  $H$ and inserted in the Hele Shaw cell. The mylar is tightly fitted so that
no flow is observable between its surface and  the walls of the cell.  The  height $\Delta W$ of the step parallel to the $y$ axis is $30\, \mathrm{mm}$ and it is located in the middle of the cell; the width $W(x)$ of the flow section in the cell is then   $70\, \mathrm{mm}$ in the inlet section and is reduced to $40\, \mathrm{mm}$ at the outlet as shown in the figure. The axis $y = 0$ corresponds to the boundary of the mylar sheet at the inlet.

Using a continuous rather than an abrupt variation of the height of the step boundary with the distance first minimizes unwanted solid-solid interactions between the fiber and this boundary. Moreover, a continuous variation of the width of the flow channel with distance is closer to the variation encountered in a real fractured rock. Finally, putting uniform flow sections before and after the step allows one to detect possible irreversible rotations and deformations of the fiber. 

A gear pump produces a downward flow; two Newtonian fluids are used: pure (MiliQ) water, of density 
$\rho = 997 \pm 1\, \mathrm{kg.m^{-3}}$ and a viscosity $\mu \approx 1 \, \mathrm{mPa.s}$, 
and a solution of $70 \%$ glycerol by weight in water,  of density $\rho = 1178 \pm  1\, \mathrm{kg.m^{-3}}$ and viscosity  $\mu \approx 15 \, \mathrm{mPa.s}$ ($27\mathrm{^o C}$).
The flow rate $Q$ is in the range $10^4 \le  Q  \le 2.3 \times 10^4 \, \mathrm{mm^3.s^{-1}}$, which  corresponds to a characteristic  velocity  $95 \le U \le 230\ \mathrm{mm.s^{-1}}$; in this work, $U$ is the maximum of the Poiseuille velocity profile (half way between the walls) in the  inlet parallel flow section upstream of the step. Downstream of the step, the velocity is higher by a factor $1.75$. For the water-glycerol solution, the corresponding Reynolds number $Re = UH/\nu$ is in the range $16  \le Re = UH/\nu \le 40$. For water, one has respectively $210 \le Re \le 500$.

The flexible silicone fibers are  made using a technique developed in the laboratory. Silicone is forced out of a syringe at an adjustable flow rate by means of a syringe pump. A second pump with a plate attached to the top is placed below the end of the first one so that the plate moves perpendicular to the flow. The injection leaves on the top of the plate  thin flexible silicone  fibers  of constant diameter.  The respective values of the speeds of the silicone at the outlet of the syringe pump and of the moving plate are very important. In order to obtain  fibers of constant diameter, the velocity 
of the outflow must match approximately that of the plate. If the speeds are not well matched, either the stream
of silicone buckles or it breaks~\cite{Habibi2007}.
Fibers of different diameters can be made by using nozzles of different sizes: in the present work, nozzles of diameter  $300-600\, \mathrm{\mu m}$  were used. The average  $D$ of the diameter over its orientation in the section and the fiber length is constant within $\pm 0.015\, \mathrm{mm}$ from a fiber to another; the diameter is constant along the length for a given orientation  within $\pm 0.05\, \mathrm{mm}$ ($D = 0.49\, \mathrm{mm}$) and  $\pm 0.025\, \mathrm{mm}$ ($D = 0.49\, \mathrm{mm}$). In a given section, the variation of the diameter with its orientation is of the order of  $\pm 0.15\, \mathrm{mm}$ for $D = 0.91\, \mathrm{mm}$ and $\pm 0.02\, \mathrm{mm}$
 for $D = 0.49\, \mathrm{mm}$. Large fibers become indeed elliptical because they sag under gravity before they polymerize: this effect is prevented by surface tension for $D \lesssim 0.7\, \mathrm{mm}$. 

The lengths of the fibers are  $L_f = 15$, $30$ and $60\, \mathrm{mm}$ and, although they are lighter than water ($\rho_f = 796\, \mathrm{kg.m^{-3}}$), the tests showed that gravity has  little effect on the experimental results. In order to evaluate the stability of their properties, 
the silicone fibers were left overnight in different fluids  (water, and water glycerol solutions): no swelling occurred.

In the interpretation of the present experiments,  the bending stiffness $E\, I$ of the fiber  plays a major role. In order to measure $E\, I$, the fiber is  allowed to bend under its own weight with one of its ends clamped in a  chuck-like device which keeps it locally horizontal; the fiber is placed in front of a dark background in order to achieve a high optical contrast. 
The process is repeated for several  fiber  lengths:   $5 \le L_f \le 80\ \mathrm{mm}$. The values of  $\delta x$, {\it i.e.} the vertical deflection of the tip of the fiber with respect to its position or a zero external load, and of  $L_f$ are determined  with an uncertainty of $\pm 20 \, \mathrm{\mu m}$ by means of an image analysis technique developed by Semin et al.~\cite{Semin2011}. In the present case of a rod bent under its own weight and clamped at one end, the theoretical expression of $\delta x$ is $m_L\, g L_f^4 / (8 E\,I)$ where  $m_L\, g$ is the force per unit length ({\it i.e.} the weight per unit length). The data were analyzed by plotting the deflection $\delta x$ as a function of the fourth power $L^4$ of the length.  The relationship between these two quantities is linear for $\delta x/L\lesssim 0.1$; this is the limit of the small deflection approximation. The slope of this plot gives the value of  $E\, I$ (See Tab.\ref{tab:fibras}). 

\begin{table}[h] 
\begin{center}
\begin{tabular}{|c|c|c|c|c|} \hline
Fiber & Material & $D$ $\mathrm{(mm)}$ & $\rho$ $\mathrm{(kg.m^{-3})}$ & $E\,I \times 10^{-8} \mathrm{(N.m^2)}$ \\  \hline
$F1$ & Silicone & $0.49 \pm 0.015$ & $0.8$ &  $0.28 \pm 0.04$\\ 
$F2$ & Silicone & $0.91 \pm 0.015$ & $0.8$ &  $0.37 \pm 0.04$ \\ 
\hline
\end{tabular}
\end{center}
\caption {Physical and mechanical characteristics of the fibers ($D$, $\rho$ and $EI$ are respectively the diameter (averaged 
over the length of the fibers and the orientation in a section), the
density and the bending stiffness of the fiber).}    
\label{tab:fibras}
\end{table}

Experimentally, after  the flow is  established, the fibers are  inserted into the upper inlet with one of the ends held by pliers. 
Three insertion points corresponding to different distances $y_{inlet}$ from the boundary of the step at the inlet have been selected (see Fig.~\ref{fig:distancia_escalon}a). 
Once the fiber is aligned with the flow, it is released and digital images of its subsequent motion in the cell
are recorded.
In order to improve the contrast, the model is illuminated from behind by a plane light panel, and images are 
obtained by means of a digital camera at a  rate of $30\, \mathrm{fps}$ and with a resolution of 
$1024 \times 768$ pixels (with $1 \mathrm{pixel} = 0.22\, \mathrm{mm}$). 
Each picture is then processed digitally in order to determine the instantaneous profile of the fibers.

The velocity of the fiber is determined from the displacement of its geometrical center  between 
two consecutive pictures.
The deformation of the fiber is characterized quantitatively 
by adjusting locally the profile by a polynomial function in order to determine the local slope and the curvature.
Only experiments in which fibers are relatively straight and parallel to the flow before reaching the step were considered 
in the analysis.

In parallel with these experiments, three dimensional numerical simulations solving the full Navier-Stokes equation
by means of the FLUENT\texttrademark  \, package have also been carried out.  These simulations provide both the
components of the velocity and their derivatives with respect to the coordinates $x$, $y$ and $z$.  The curvature of the 
streamlines and the derivative $\partial v/\partial n$ of the velocity  with respect to distance perpendicular  to these
lines is then computed by combining the values of these derivatives. 
\section{Experimental results}
\label{sec:B}
\subsection{Characterization of the flow field}
\label{sec:B-1}
\begin{figure}[htbp]
\includegraphics[width=\WL]{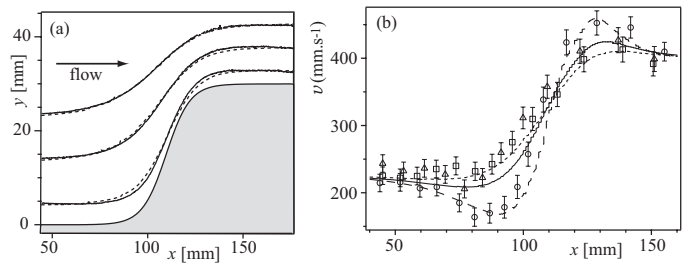}
\caption{(a)  Visualization of three streamlines  by the injection of a dye (continuous lines). Gray region: step obstacle; dotted lines: streamlines computed numerically for the same injection points. (b) Numerical (lines) and experimental (symbols) variation of the fluid velocity $v$ along  streamlines  corresponding to   three injection points located at $x = 0$ and  $y_{inlet} = 4.5 \pm 1.5\, \mathrm{mm}$ (($\circ$), dashed line), $13.5 \pm 2.0\, \mathrm{mm}$ ($\triangle$, continuous line) and $21.5 \pm 2.0\, \mathrm{mm}$ ($\square$, dotted line). The characteristic velocity $U$ upstream of the obstacle is equal to: $230\ \mathrm{mm.s^{-1}}$. Note that the horizontal and vertical length scales in Fig. a are different.}
\label{fig:distancia_escalon} 
\end{figure}
Prior to the injection of the fibers, we have studied the flow velocity variations along the streamlines starting at each of the three selected injection points.
Experimentally, a stationary flow is first established in the cell  and a stream of dye is then injected at each location 
(Fig.~\ref{fig:distancia_escalon}a). 
The progression of the dye  is recorded and analyzed using the procedure described in the previous section. 
The tip of the dye streak moves at the maximum of the local Poiseuille profile in the gap: this allows one to measure quantitatively the variation of this maximum along the corresponding streamline.  The results 
obtained from these measurements are then compared to those of the $3D$ numerical simulations.

Figs.~\ref{fig:distancia_escalon}a-b show that  the experimental and numerical results are in good agreement for both the geometry of the streamlines (a) and the variations of the fluid velocity along them (b).  For the most distant streamlines (dotted line and squares in Fig.~\ref{fig:distancia_escalon}), the velocity increases continuously with distance along $x$ in direct relation with  the decrease  of the flow section. The results are different for the  two other streamlines: the velocity first drops as the fluid reaches the vicinity of the step, then it increases sharply, reaches a maximum right after the step and decreases finally towards a constant value. Localized buckling may therefore occur in these small regions of decreasing velocity.

Fig.~\ref{fig:distancia_escalon} corresponds to an upstream velocity $U = 230\, \mathrm{mm.s^{-1}}$ 
(in the mid plane between the walls): these numerical simulations were also performed for a velocity $10$ times lower. As expected from the Hele Shaw approximation, the streamlines of interest at both velocities (and their curvature) are identical within experimental error while the  velocity components are everywhere proportional to $U$.
\begin{figure}[htbp]
\includegraphics[width=\WL]{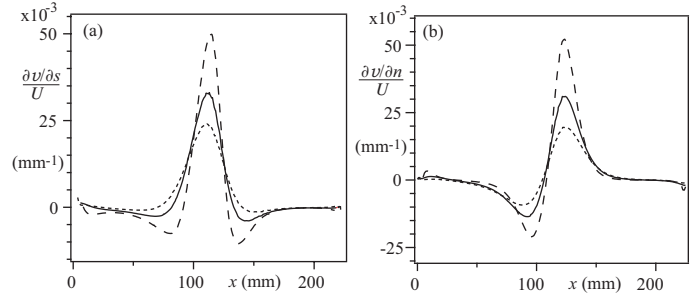}
\caption{Variations as a function of the distance $x$ of the normalized parallel and transverse components $(\partial v/\partial s)/U$ (a) and $(\partial v/\partial n)/U$ (b) of the gradient of flow velocity $v$ computed numerically for the same streamlines as in Fig.~\ref{fig:distancia_escalon}b (the different types of lines have the same meaning as in this latter figure).}
\label{fig:gradients_vitesse} 
\end{figure}

Actually, the deformations of the fibers are not due to the $v$ velocity itself but to its spatial variations (see Section~\ref{sec:B-2b}). 
The variations with the distance $x$ of the normalized gradient components $(\partial v/\partial s)/U$ and $(\partial v/\partial n)/U$ parallel and transverse to the velocity are plotted in Figs.~\ref{fig:gradients_vitesse}a-b for the same streamlines as in
 Fig.~\ref{fig:distancia_escalon}b. The parallel component  $(\partial v/\partial s)/U$  has its highest positive value (inducing
 an extensional force) in the region of maximum slope of the streamlines (Fig.~\ref{fig:gradients_vitesse}a); 
 it is instead negative  (and produces   a compressional force which may induce buckling) at the very beginning and end 
 of the steps. The transverse component $(\partial v/\partial n)/U$ 
 which  will be seen to induce  deformations of the fiber (see Sec.~\ref{sec:B-2}) goes instead to zero when the slope
 of the streamlines is highest; ($\partial v/\partial n)/U$ reaches  its maximum positive and negative values
 in the regions of maximum curvature (Fig.~\ref{fig:gradients_vitesse}b). The width of these extrema is of the order of 
$30\, \mathrm{mm}$.
 
Quantitatively, Tab.~\ref{tab:tab2} lists the minimal (negative) and maximal values (as determined from the numerical simulations) of the curvature of the line together with the transverse normalized velocity gradient $(\partial v/\partial n)/U$ at the same locations; as mentioned above, these values are independent of $U$. At the distances $x = 90$ and $130\, \mathrm{mm}$  corresponding to the minimal and maximal velocities along the streamline closest to the obstacle, one finds respectively: $(\partial v/\partial n) \sim 0.015 \, U$ and $0.04  \, U$.  
The experimental values are  very close to the numerical ones.

\begin{table}[h]
\begin{center}
\begin{tabular}{|c|c|c|c|c|} \hline
 Init. location & \multicolumn{2}{c|}{Max. curvature (upstr.)} & \multicolumn{2}{c|}{Min. curvature (downstr.)}  \\ \hline
 $y_{inlet}$ & $\kappa^{max}_{s}$ & $(1/U) (\partial v/\partial n)$ & $\kappa^{min}_{s} $ & $(1/U) (\partial v/\partial n)$ \\  \hline
  4.5 & $0.0144$ & $-0.014$ & $-0.0212$ & $0.039$\\ 
 13.5  & $0.0109$ & $-0.006$ & $-0.011$ & $0.0222$ \\ 
 21.5 & $0.0077$ & $-0.0034$ & $-0.0066$ & $0.011$ \\  \hline
\end{tabular}
\end{center}
\caption{Transverse coordinate $y_{inlet}$ (in mm) at inlet, maximum ($\kappa^{max}_{s}$) and minimum ($\kappa^{min}_{s}$) curvatures and velocity  gradient $(1/U) (\partial v/\partial n)$ in mm$^{-1}$ at the corresponding locations  for the three streamlines shown in Fig.~\ref{fig:distancia_escalon}.}
\label{tab:tab2}
\end{table}
\subsection{Mean velocity and orientation of the fiber}
\label{sec:B-1bis}
\begin{figure}[htbp]
\includegraphics[width=\WL]{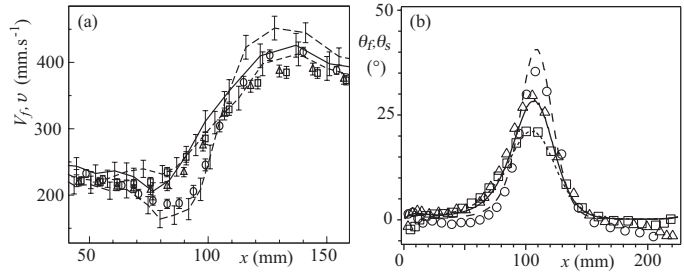}
\caption{(a) Experimental variations as a function of the distance $x$ of the flow velocity $v$ measured by the injection of dye (lines) and of the velocity $V_f$ of the center of mass of a $30\,\mathrm{mm}$ long fiber of diameter $D=0.49\ mm$ (symbols) along  streamlines corresponding to the three same injection points as in Fig.~\ref{fig:distancia_escalon}. (b) Variation as a function of the distance $x$ of the local angles with the axis $x$ of the flow velocity (lines) and of the fiber (symbols). Fluid velocity upstream of the obstacles: 
$U = 230\, \mathrm{mm.s^{-1}}$.  The meaning of the symbols and of the different types of lines is  the same as 
in Fig.~\ref{fig:distancia_escalon}b.}
\label{fig:velocity_orientation} 
\end{figure}
We have studied the velocity $V_f$ of the center of mass of fibers  of diameters $D=0.49$ and $0.91~\mathrm{mm}$ corresponding to confinement ratios $D/H = 0.22$ and $0.41$ and initially straight and parallel to the flow.
In the parallel flow region upstream of the step where  $v = U$, one finds that: $V_{f}=(0.97 \pm 0.05) U$ and $V_{f} = (0.87  \pm  0.1) U$. The ratio $V_f/U$ only depends on the diameter of the fibers and not of their length. Moreover, from Fig.~\ref{fig:velocity_orientation}a, the velocity $V_f$ remains close to $v$ at all distances along the streamline and not only in its parallel part. These results agree with an experimental and numerical study by other authors~\cite{Berthet2013}  
of the transport of rigid fibers  by a Poiseuille  flow  between  parallel flat walls. For the same ratios $D/H = 0.22$ and $0.41$,
these authors predict that $V_f$ is respectively only $4\%$ and $10\%$ lower than $v$. $V_f$ is, in addition, the same for
fibers parallel and perpendicular to the flow.
This small difference between $V_f$
and $v$, particularly for $D/H = 0.22$, implies that the friction forces
due to the relative motion of the fiber and the walls do not play a major part in the phenomenon.
The velocity of the fluid in the center part of the cell gap is, therefore, used as the relevant velocity of the problem.

As the fiber moves along the streamline, it becomes tilted with respect to the vertical axis $x$ as it  moves over the step: 
the variations of the tilt
angle $\theta_f$ with the distance $x$ for the three streamlines are plotted in Fig.~\ref{fig:velocity_orientation}, together with those of the
local tilt angle $\theta_s$ of
the streamline at the same distance. The variations of $\theta_f$ and $\theta_s$ are very similar, particularly for the two streamlines
 farthest from the contour of the step. For that closest to the contour, instead,
the maximum tilt angle of the fiber is slightly lower than that of the streamline; moreover, $\theta_f$ does not revert
exactly to zero after the step. This suggests that a part of the fiber moves into the layer close to the contour  where
the Hele Shaw approximation of zero 2D vorticity in the $(x,y)$ plane is not valid any more, resulting in a residual rotation. When the fiber happens to be injected particularly close to the obstacle, this rotation may be so large that the fiber touches the obstacle: such experiments are not included in the present interpretations.
\subsection{Deformations  of fibers of constant length}
\label{sec:B-2}
\begin{figure}[htbp]
\includegraphics[width=\WL]{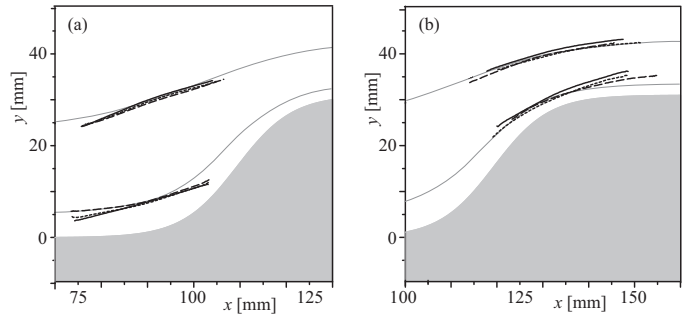}
\caption{Fibers ($L_f = 30\, \mathrm{mm}$, $D = 0.49\, \mathrm{mm}$) observed at  two particular locations 
upstream (left) and downstream (right) of the step for three upstream velocities: $U = 95 \, \mathrm{mm.s^{-1}}$ (black), $160 \, \mathrm{mm.s^{-1}}$ (light gray) and $230 \, \mathrm{mm.s^{-1}}$ (gray). Streamlines correspond to 
$y_{inlet} = 4.5\, \mathrm{mm}$ (dashed line) and $y_{inlet} =21.5 \, \mathrm{mm}$ (dotted line). 
Fluid viscosity: $\mu = 15\, \mathrm{mPa.s}$.}
\label{fig:Vitesse} 
\end{figure}
In this section, we compare  the deformations of fibers of same length $L_f = 30\,\mathrm{mm}$ at  different distances 
$x$ on a given streamline and for three different values of the mean flow velocity.
Fig.~\ref{fig:Vitesse} displays superimposed images of  fibers of same diameter $D = 0.49\, \mathrm{mm}$ obtained at these different velocities in 
the  regions of maximum positive (a) and negative (b) curvature of two different streamlines (these regions are 
located respectively on the upstream and downstream sides of the step). In case (b), increasing the  
fluid velocity  strongly increases  the curvature of the fiber which  becomes of the order of that of the streamline:
this is particularly visible on the streamline closest to the  contour of the step. For the other, the curvature is always similar to
  that (weaker) of the other streamline. In case (a), the effect of the velocity
is still clear close to the contour but much weaker on the other streamline.

\begin{figure}[htbp]
\begin{center}
\includegraphics[width=\W]{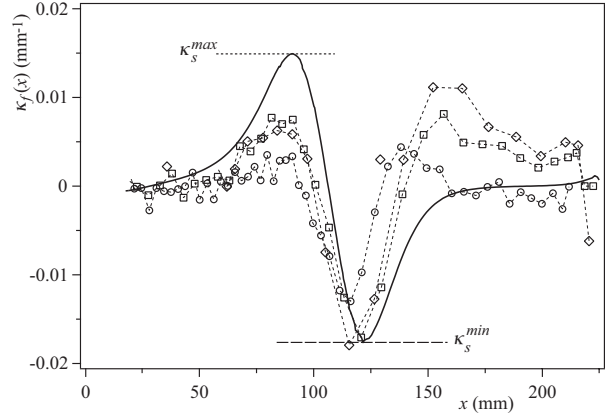}
\caption{Curvature $\kappa_f  \mathrm{(mm^{-1})}$ of a fiber ($D = 0.49\, \mathrm{mm}$, $L_f = 30\, \mathrm{mm}$) 
as a function of the coordinate $x$ of its geometrical center for an initial location $y_{inlet} = 13.5\, \mathrm{mm}$ and
 for  flow velocities:  ($\bigcirc$) $U = 95$ , ($\square$) $168$  and  ($\diamond$)
$230 \, \mathrm{mm.s^{-1}}$. Continuous line: curvature of the corresponding streamline at the same distance.
 Fluid viscosity: $\mu = 15\, \mathrm{mPa.s}$.  $\kappa^{max}_{s}$,  $\kappa^{min}_{s}$:  maximum and minimum 
 curvatures of the streamlines.}
\label{fig:Curvature-vs-position} 
\end{center}
\end{figure}
These informations are complemented by Fig.~\ref{fig:Curvature-vs-position} displaying the variation with distance of
 the  curvature $\kappa_f$ of the same fiber as it moves along the streamline corresponding
 to $y_{inlet} = 13.5\, \mathrm{mm}$. This variation is compared for different velocities $U$   to that of 
 the local curvature $\kappa_s$ of the streamline: $\kappa_f$ reaches extremal  values at
  distances $x$ close to those corresponding to the extrema $\kappa_s^{max}$ and $\kappa_s^{min}$. 
  The minimum of $\kappa_f$ downstream
 of the obstacle is of the order of the minimum value of $\kappa_s$ at all velocities while it is significantly lower and
 increases with $U$ for the maximum upstream. Up to the end of the step,  the curvatures of the
 fiber and of the streamline follow a similar trend of variation with the distance $x$, even though the amplitudes 
are generally different. Farther downstream and at the highest velocity,  the curvature of the fiber displays a second 
 maximum while that of the streamline returns to zero. This may be due to buckling induced by the locally 
 negative value of the derivative $\partial v/\partial s$ shown in Fig.~\ref{fig:gradients_vitesse}a and resulting in a
  compressional  stress on the fiber just downstream of the step.
  Figs.~\ref{fig:Vitesse} and \ref{fig:Curvature-vs-position} both show
  therefore that  the absolute curvature of the fiber increases with the local  curvature of the streamlines 
  but depends also  of the local flow.

We study now quantitatively the values of these extremal curvatures at
different flow rates. Since the flow is laminar, the curvature of the fibers corresponds to a   balance 
between  elastic and viscous  forces: the latter   are induced by {\it local} velocity differences between 
different parts of the fiber and the fluid. These differences are small compared to the characteristic flow
velocity $U$ since the  {\it global} fiber velocity is close to that of the fluid: the corresponding Reynolds number
 is also small and the viscous  stresses are proportional to these velocity contrasts. The relation of the latter  to 
 the spatial variations of the velocity and the deformation of the fiber and the streamlines is discussed
 in Sec.~\ref{sec:B-2b} but, as a first step, $\mu\, \partial{v}/\partial{n}$ will be chosen as the  reference viscous stress: 
 $\partial{v}/\partial{n}$ is indeed strongly correlated to the  variations of the curvature, 
as shown by Figs.~\ref{fig:gradients_vitesse} and \ref{fig:Curvature-vs-position}.
 
\begin{figure}[htbp]
\includegraphics[width=\W]{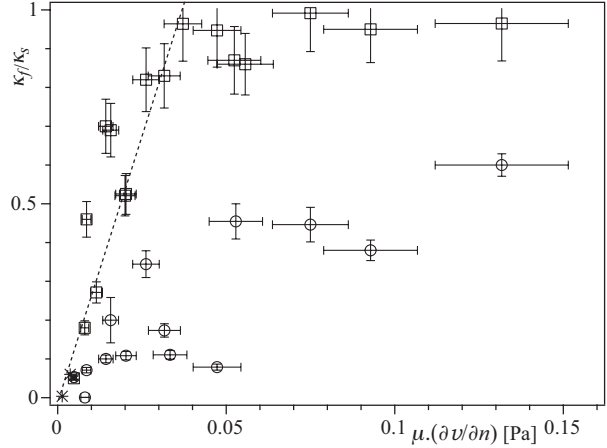}
\caption{Dimensionless curvature, $\kappa_f/\kappa_s$, of fibers of length  $L_f = 30\, \mathrm{mm}$ as a
function of the viscous stress $\mu (\partial v/\partial n)$. Diameter $D = 0.49\, \mathrm{mm}$; $(\square)$: 
viscosity $\mu=15\, \mathrm{mPa.s}$, ($\mathrlap{+}{\times}$):  $\mu = 1\, \mathrm{mPa.s}$. 
$(\circ)$: $D = 0.91\, \mathrm{mm}$,   $\mu=15\, \mathrm{mPa.s}$ .}
\label{fig:ForceVisqueuse} 
\end{figure}
Fig.~\ref{fig:ForceVisqueuse} displays variations of $\kappa_f/\kappa_s$  as a function of the viscous stress 
$\mu\, \partial{v}/\partial{n}$ 
($\kappa_f/\kappa_s$ is used as the parameter of interest since, from Fig.~\ref{fig:Curvature-vs-position},
 one has always $\kappa_f \le \kappa_s$).
Significant curvatures $\kappa_f/\kappa_s$ are only observed for the water-glycerol solution (viscous stresses are too low 
for water ($\mathrlap{+}{\times}$ symbols)). As expected, for a given $\mu\, \partial{v}/\partial{n}$, the curvatures are also highest for the most flexible fiber ($D = 0.49\,\mathrm{mm}$). In this latter case ($\square$ symbols), $\kappa_f/\kappa_s$ increases at first linearly with $\mu\, \partial{v}/\partial{n}$ and tends then toward a limit of the order of $1$ (the fiber coincides locally with
 the streamline). This limit is not reached for the less flexible fibers ($\circ$ symbols).
\subsection{Characteristic dimensionless  number}
\label{sec:B-2b}
In order to collapse the data corresponding to different fibers  onto a single master
curve, it is  necessary to replace as the horizontal scale $\mu (\partial v/\partial n)$ by 
a combination including, in addition, the diameter and the 
length of the cylinder and the stiffness of its material. 
As mentioned above, the  degree of bending of the fiber reflects a balance between the
 viscous and elastic forces upon it. 
The value of the diameter influence both of them:
 
\begin{itemize} 
\item The curvature $\kappa_f$ of a fiber of length $L_f$ submitted to a force $q$ per unit length
 scales as $q L_f^2/(E\,I)$: $E\,I$ is the bending modulus with $I = \pi D^4/64$. The force necessary
 to achieve a given curvature scales therefore as $D^{4}$.
\item  In the present case, $q$ is a  viscous stress associated to the local relative velocity 
$q$ also depends on $D$ through the confinement ratio $D/H$: Increasing
 $D/H$ results indeed in a stronger blockage of the flow normal to the fiber and, thus, enhances the normal 
 viscous stress. This confinement effect is taken into account  by introducing a  factor
$\lambda_p^{\perp}(D/H)$  in the proportionality relation between the viscous stress and both the 
relative velocity of the fiber and the fluid and the viscosity of the latter. 
The influence of $D/H$ on the force on fixed circular cylinders in a similar flow geometry has been 
studied in  Refs.~\onlinecite{Richou2004,Richou2005,Semin2009}. For the ratios $D/H = 0.22$ 
and $0.41$ corresponding to $D = 0.49$ and $0.91\, \mathrm{mm}$, these studies provide respective values  
$\lambda_p^{\perp}=30$ and $73$.
\end{itemize} 

As suggested  in the previous section, the viscous forces are proportional to the local relative velocities of
the different parts of the cylinders and the fluid in the regions where the streamlines are curved.
These relative velocities may be estimated by multiplying $\partial v/\partial n$ by the local distance between
the ends of the   fiber (of curvature $\kappa_f$) and a streamline of curvature $\kappa_s$: this distance
is then of the order of $(\kappa_s - \kappa_f) L_f^2$ so that we take: 
$q \propto \lambda_p^{\perp} \, \mu\, (\kappa_s - \kappa_f)  L_f^2 (\partial v/\partial n)$.
The  factor $\lambda_p^{\perp}$ takes into account the confinement of the flow
Using this expression in the estimation of the curvature $\kappa_f$  leads to:

\begin{equation}\label{eq:spermf}
\frac{\kappa_f}{\kappa_s - \kappa_f}  = C S_{p} =  C \frac{\lambda_p^{\perp}\, \mu\,  L_f^4 (\partial v/\partial n)}{E\,I}.
 \end{equation}
The right hand side of the equation is the Sperm number $S_p$ defined by Eq.~(\ref{eq:sperm}) with
the additional factor $\lambda_p^{\perp}$.
\begin{figure}[htbp]
\includegraphics[width=\W]{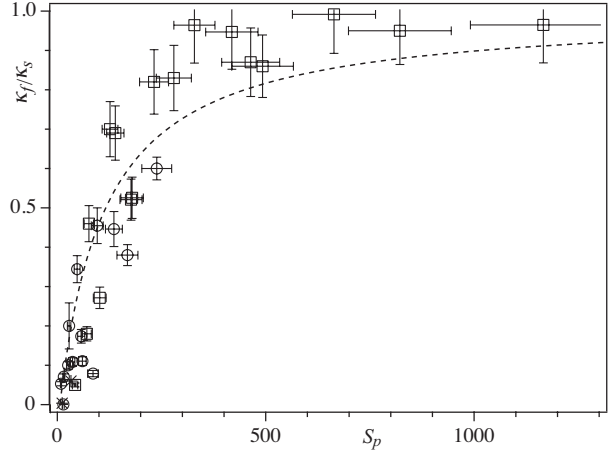}
\caption{Dimensionless curvature, $\kappa_f/\kappa_s$ of fibers of constant length: $L_f = 30\ \mathrm{mm}$ as a function of the Sperm number  $S_p = \lambda_p^\perp \mu (\partial v/\partial n)  L_f^4/EI$. Fiber diameter: $D = 0.49 \, \mathrm{mm}$
 ($\square$, $\mathrlap{+}{\times}$),  $D=0.91\, \mathrm{mm}$ ($\circ$); fluid viscosity: $\mu=15\, \mathrm{mPa.s}$ ($\circ$, $\square$),   
 $\mu=1\ mPa.s$ ($\mathrlap{+}{\times}$). Dashed line: predictions from Eq.~(\ref{eq:spermf}) with
  $C =1/110$.}
\label{fig:S-vs-Sp-L30} 
\end{figure}

Fig.~\ref{fig:S-vs-Sp-L30} shows the variation of $\kappa_f/\kappa_s$  as a function of $S_p$. This time, the data corresponding
to the two different fiber diameters collapse  onto a same global trend.  
At low values of $S_p \ll 100$, the ratio $\kappa_f/\kappa_s$ increases steeply with $S_p$ ($\kappa_f/\kappa_s \simeq S_p/110$
for the dashed line).
At large values of $S_p \gtrsim 500$, $\kappa_f/\kappa_s$ becomes of the order of $1$ and the profile of the fiber follows the
streamlines. The dashed line corresponds to the best fit of these experimental data by Eq.~(\ref{eq:spermf}) which is obtained 
for $C = 1/110$ so that $\kappa_f/\kappa_s = S_p/(110 + S_p)$: the fitted curve follows well the trend of the experimental data.

Two of the data points plotted in Fig.~\ref{fig:S-vs-Sp-L30} correspond to the  lower viscosity $\mu=1\, \mathrm{mPa.s}$.  
Like in Fig.~\ref{fig:ForceVisqueuse} (and for the same reasons), the corresponding values of $\kappa_f/\kappa_s$ are 
significantly lower.
\subsection{Influence of the length of the fiber}
\label{Sec:length}
\begin{figure}[htbp]
\includegraphics[width=\WL]{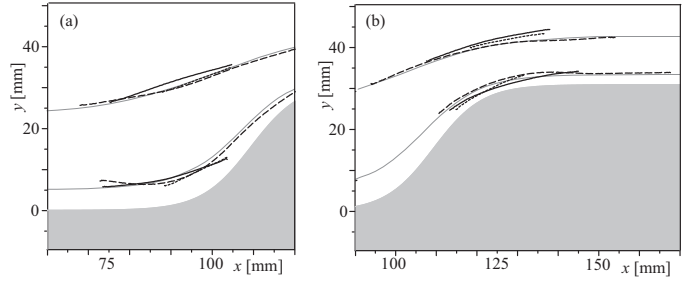}
\caption{Superimposition of fibers of different lengths  $L_f \,=\, 15$ (dotted lines), $30$ (continuous lines) and $60\ \mathrm{mm}$ (dashed lines)  over two streamlines (thin grey lines) at  two  locations  upstream (left) and downstream (right) of the step.  Fluid velocity: $U = 230 \, \mathrm{mm.s^{-1}}$; transverse locations of injection point: $y_{inlet} = 4.5$ and $21.5\, \mathrm{mm}$}
\label{fig:Length} 
\end{figure}

All the experiments described above were performed with fibers of length $L_f = 30 \, \mathrm{mm}$.
Further measurements of the curvature and longitudinal velocity  have been performed both for shorter  and
 longer fibers  (Fig.~\ref{fig:Length}): increasing the length $L_f$ increases indeed the  normal viscous stress
  ($\propto L_f^2$) while  it reduces the elastic force needed for a given bending ($\propto L_f^{-2}$).
In agreement with Eq.~(\ref{eq:spermf}), $\kappa_f/\kappa_s$  should then  scale like $L_f^4$ while it is $\ll 1$. 

Fig.~\ref{fig:Length} displays the shapes of fibers of different lengths  carried over the obstacle by a  flow of given velocity and viscosity.  While the curvature decreases clearly as  $L_f$ drops from $30$ to $15\ \mathrm{mm}$,  its value for the longest fiber ($L_f = 60\ \mathrm{mm}$) is similar to that for $L_f = 30\ \mathrm{mm}$. As seen in Fig.~\ref{fig:Curvature-vs-position}, bending is indeed localized to the regions of maximum curvature of the streamlines. Therefore,  for $L_f = 60\ mm$, only a fraction of the length of the fiber is  exposed to a large normal viscous stress: this exposed length is of the order of the width $\Delta W  \simeq 30\ \mathrm{mm}$ of each extremum of $\partial v/\partial n$ in the curves of Fig.~\ref{fig:gradients_vitesse}b and is independent of $L_f$. For long fibers ($L_f \gtrsim \Delta W$), the relevant length to be used in the expression of $S_p$ is therefore  $\Delta W$ instead of $L_f$. For this reason, we replace  the expression of 
$S_p$ from Eq.~(\ref{eq:spermf}) by the more general expression: 
\begin{equation}\label{eq:spermfxi}
S_p =  \frac{\lambda_p^\perp\, \mu\,  \xi^4 (\partial v/\partial n)}{E\,I},
 \end{equation}
in which $\xi = L_f$ for $L_f \le \Delta W$ and $\xi = \Delta W$ for $L_f \ge  \Delta W$.
\begin{figure}[htbp]
\includegraphics[width=\W]{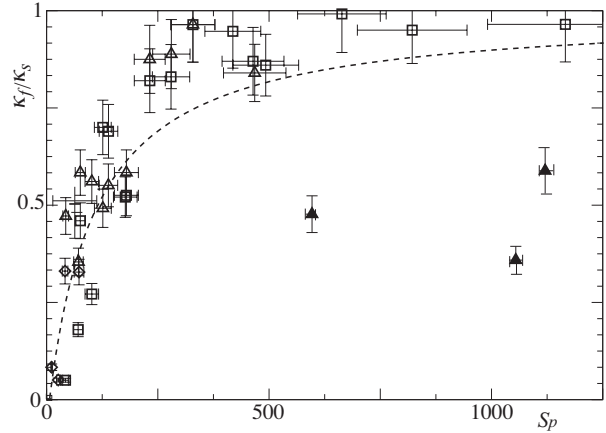}
\caption{Dimensionless curvature $\kappa_f/\kappa_s$ as a  function of $S_p$ for different lengths of the fibers: $L_f =15\, \mathrm{mm}$ (($\lozenge$)  
$D = 0.49\, \mathrm{mm}$), $L_f = 30\, \mathrm{mm}$ (($\bigcirc$) $D = 0.49\, \mathrm{mm}$, ($\square$) $D = 0.91\, \mathrm{mm}$) and $
L_f = 60\ \mathrm{mm}$ ($D = 0.91\, \mathrm{mm}$) with $\Delta W$ ($\triangle$) and $L_f$ ($\blacktriangle$) as the characteristic length in the definition of $Sp$. Fluid viscosity: $\mu=15\, \mathrm{mPa.s}$.  Dashed line: predictions from Eq.~(\ref{eq:spermf}) with a proportionality factor $1/110$.}
\label{fig:S-vs-Sp} 
\end{figure}
The validity of these assumptions is tested  in Fig.~\ref{fig:S-vs-Sp} displaying  the variation of $\kappa_f/\kappa_s$ with  $S_p$ for fibers 
of different lengths and diameters. For the two shortest fibers   $L_f  = 15$ and $L_f = 30\ \mathrm{mm}$, one has $L_f \le \Delta W$ so that both definitions of $S_p$ are valid;  for  $L_f =60\, \mathrm{mm}$, a better collapse is clearly obtained by using the 
definition of $Sp$ from Eq.~(\ref{eq:spermfxi}) ($\triangle$) than from  Eq.~(\ref{eq:spermf}).
\section{Discussion and conclusion}
The present experiments  identified the  parameters controlling the
motion and deformation  of a flexible fiber in a confined Hele-Shaw  flow where a step-like obstacle induces  variations of  the magnitude and orientation of the velocity.
 
Both the trajectory of the center of mass of the fiber and its orientation follow globally those of the corresponding
streamline. This contrasts with pinned flexible fibers or fibers placed in a Poiseuille flow, for which the fibers may cross the streamlines. Moreover, the velocity of the fibers is proportional to the fluid velocity  at their center of mass  with a constant only depending of the geometrical confinement. 
The variations of the curvature  $\kappa_f$  of the fiber also follow closely those $\kappa_s$ of the streamlines but, generally, with a smaller amplitude. 

Quantitatively, a key result of the present study is that the normalized curvature $\kappa_f/\kappa_s$ of the fibers  
is only a function  of a combination  $S_p$ (the Sperm number) of  different parameters of the flow given 
by Eq.~(\ref{eq:spermfxi}). Physically, relation (\ref{eq:spermf}) between $\kappa_f/\kappa_s$ and $S_p$ reflects the balance
between  viscous forces applied by the flowing fluid  on the fiber and  elastic forces opposing the resulting deformations. The viscous forces are proportional   to  the
transverse velocity gradient,   to the local distance between the fiber and the streamline and to the fluid viscosity.
The elastic restoring forces, per unit length, are proportional   to the curvature of the fiber,  to its bending stiffness $E\, I \propto D^4$ and  vary as $\xi^{-2}$ with its effective length $\xi$ for a given curvature: $\xi$ coincides with the length $L_f$ only for short fibers. 
  For longer ones, $\xi$ must be taken equal in the expression of $S_p$ to a characteristic length of the variation of the flow. 
  Here, if $L_f \ge \Delta W$, $\xi$ is equal to the width $\Delta W$ of the region where the component $\partial v/\partial n$ of the gradient  of the  velocity is large (Fig.~\ref{fig:gradients_vitesse}b); parts of the fibers 
  outside this region do not  influence  bending.
  
 In contrast with many previous studies of the dynamics of flexible fibers, one deals here with a confined configuration of a fiber between two parallel walls: this keeps the motion of the fiber bi-dimensional  and enhances the viscous forces due to the partial 
  blockage of the flow: this effect is taken care of by introducing a dimensionless parameter $\lambda_p^\perp$ depending on the
  diameter. The diameter influences however bending mainly through the $D^4$ factor in $EI$ which increases 
  the rigidity considerably and reduces bending for fibers of larger diameter.
 
The  present study can be applied to the transport of flexible  fibers in fractures containing obstacles occupying a part of the area:
the quantitative information we obtained can be transposed to estimate  the influence of the length and diameter of the fibers,
of the fracture gap and the size in the fracture plane of the obstacles  (modeled here as step reducing the width of an Hele-Shaw cell).

In the present flow geometry, the velocity of the flow increases with distance which keeps the fiber under tension, except  in  localized regions. It will be important to investigate the opposite case in which buckling may, for instance,
occur. 
The  discussion presented above is largely based on a local balance between elastic and viscous forces. Non local effects 
might however occur for  fibers submitted to a sequence of variations of curvatures of the streamlines over
distances smaller than their length. Interactions between the different parts of the  fiber may then be significant: this may have an important  influence on transport in  porous media or micro-fluidic 
circuits.
\begin{acknowledgments} 
We acknowledge support by the RTRA Triangle de la Physique
and the LIA PMF-FMF (Franco-Argentinian International Associated Laboratory
in the Physics and Mechanics of Fluids). We thank R. Pidoux for the realization
of the experimental cell. We are grateful to G. Kasperski for his help with the numerical
simulations of the flow field. 
\end{acknowledgments} 


\begin{thebibliography}{0}%
\makeatletter
\providecommand \@ifxundefined [1]{%
 \@ifx{#1\undefined}
}%
\providecommand \@ifnum [1]{%
 \ifnum #1\expandafter \@firstoftwo
 \else \expandafter \@secondoftwo
 \fi
}%
\providecommand \@ifx [1]{%
 \ifx #1\expandafter \@firstoftwo
 \else \expandafter \@secondoftwo
 \fi
}%
\providecommand \natexlab [1]{#1}%
\providecommand \enquote  [1]{``#1''}%
\providecommand \bibnamefont  [1]{#1}%
\providecommand \bibfnamefont [1]{#1}%
\providecommand \citenamefont [1]{#1}%
\providecommand \href@noop [0]{\@secondoftwo}%
\providecommand \href [0]{\begingroup \@sanitize@url \@href}%
\providecommand \@href[1]{\@@startlink{#1}\@@href}%
\providecommand \@@href[1]{\endgroup#1\@@endlink}%
\providecommand \@sanitize@url [0]{\catcode `\\12\catcode `\$12\catcode
  `\&12\catcode `\#12\catcode `\^12\catcode `\_12\catcode `\%12\relax}%
\providecommand \@@startlink[1]{}%
\providecommand \@@endlink[0]{}%
\providecommand \url  [0]{\begingroup\@sanitize@url \@url }%
\providecommand \@url [1]{\endgroup\@href {#1}{\urlprefix }}%
\providecommand \urlprefix  [0]{URL }%
\providecommand \Eprint [0]{\href }%
\providecommand \doibase [0]{http://dx.doi.org/}%
\providecommand \selectlanguage [0]{\@gobble}%
\providecommand \bibinfo  [0]{\@secondoftwo}%
\providecommand \bibfield  [0]{\@secondoftwo}%
\providecommand \translation [1]{[#1]}%
\providecommand \BibitemOpen [0]{}%
\providecommand \bibitemStop [0]{}%
\providecommand \bibitemNoStop [0]{.\EOS\space}%
\providecommand \EOS [0]{\spacefactor3000\relax}%
\providecommand \BibitemShut  [1]{\csname bibitem#1\endcsname}%
\let\auto@bib@innerbib\@empty
\end{thebibliography}%


\begin{thebibliography}{References}
\bibitem{Yasuda2004}K. Yasuda, T. Kyuto and N. Mori, ``An experimental study of flow-induced fiber orientation and concentration distributions in a concentrated suspension flow through a slit channel containing a cylinder,'' Rheol. Acta.  {\bf 43}, 137-145 (2004).
\bibitem{Tornberg2006}A-K. Tornberg, K. Gustavsson, ``A numerical method for simulations of rigid fiber suspensions,'' J. Comp. Phys. {\bf 215}, 172-196 (2006).
\bibitem{Howard1995} P.R. Howard, M.T. King, M. Morris, J-P. Feraud, G. Slusher, S. Lipari,
``Fiber/Proppant Mixtures Control Proppant Flowback in South Texas,'' SPE Annual Technical Conference and Exhibition, 22-25 October 1995, Dallas (TX), 453-454 (1995). 
\bibitem{Purcell1976}E.M. Purcell, ``Life at Low Reynolds Number,'' Am. J. Phys. {\bf 45}, 3-11 (1977).
\bibitem{Lauga2009}E. Lauga and T. R Powers, ``The hydrodynamics of swimming microorganisms,'' Rep. Prog. Phys. \textbf{72},  096601 (2009).
\bibitem{Lagomarsino2003} M.C. Lagomarsino, F. Capuani, C.P. Lowe, ``A simulation study of the dynamics of a driven filament in an Aristotelian fluid,'' J. Theor. Biol.  \textbf{224}, 215-224 (2003).
\bibitem{Lagomarsino2004} M.C. Lagomarsino, I. Pagonabarraga and C. Lowe, ``Hydrodynamic induced deformation and orientation of a microscopic elastic filament,'' Phys. Rev. Lett. \textbf{94}, 148104 (2005).
\bibitem{Chelakkot2010}R. Chelakkot, R.G. Winkler and G. Gompper, ``Migration of semiflexible polymers in microcapillary flow,'' Europhys. Lett. \textbf{91},  14001  (2010).
\bibitem{Reddig2011}S. Reddig and H. Stark, ``Cross-streamline migration of a semiflexible polymer in a pressure driven
flow,'' J. Chem. Phys. \textbf{135}, 165101 (2011).
\bibitem{Slowicka2012}A. M. Slowicka, M. L. Ekiel-Jezewska, K. Sadlej, and E. Wajnryb, ``Dynamics of fibers in a wide microchannel,''
J. Chem. Phys. \textbf{136}, 044904 (2012).
\bibitem{Rusconi2010} R. Rusconi, S. Lecuyer, L. Guglielmini and H.
Stone, ``Laminar flow around corners triggers the formation of biofilm streamers,'' 
J. R. Soc. Interface. \textbf{{7}}, 1293-1299 (2010). 
\bibitem{Dangelo2009} M. D'Angelo, B. Semin, G. Picard, M. Poitzsch, J-P. Hulin, H. Auradou, ``Single Fiber Transport in a Fracture Slit: Influence of the Wall Roughness and of the Fiber Flexibility,'' Transp. Porous Med. \textbf{84}, 389-408 (2010).
 \bibitem{Forgacs1959}O.L. Forgacs and S.G. Mason, ``Particle motions in sheared suspensions. X: Orbits of flexible threadlike particles,'' J. Coll. Int. Sci. \textbf{14}, 473-491 (1959).
\bibitem{Stockie1998} J. M. Stockie and S. I. Green ``Simulating the Motion of Flexible Pulp fibers Using the Immersed. Boundary Method,`` J. Comp. Phys., \textbf{147}, 147-165 (1998).
\bibitem{Wang2006}G. Wang, W. Yu, C.  Zhou, ``Optimization of the rod chain model to simulate the motions of a long flexible fiber in simple shear flows,'' Eur. J. Mech. B/Fluids. \textbf{25}, 337-347 (2006)
\bibitem{Lindstrom2007}S.B. Lindstr\"om and T. Uesaka, ``Simulation of the motion of flexible fibers in viscous fluid flow, '' Phys Fluids. {\bf 19}, 113307 (2007).
\bibitem{Wandersman2010} E. Wandersman, N Quennouz, M. Fermigier, A. Lindner, and O. du Roure, ``Buckled in translation,'' Soft matter. \textbf{6}, 5715-5719 (2010).
\bibitem{Young2007} Y.N Young, M. J. Shelley, ``Stretch-coil transition and transport of fibers in cellular flows,'' Phys. Rev. Lett. \textbf{99}, 056303 (2007).
\bibitem{Guglielmini2012}L. Guglielmini, A. Kushwaha, E.S. G. Shaqfeh, and H.A. Stone, ``Buckling transitions of an elastic filament in a viscous stagnation point flow,'' Phys. Fluids \textbf{24}, 123601 (2012).
\bibitem{Kantsler2012}V. Kantsler and R.E. Goldstein, "Fluctuations, dynamics, and the Stretch-Coil transition of single actin filaments in extensional flows," Phys. Rev. Lett. {\bf108}, 038103 (2012).
\bibitem{Autrusson2011}N. Autrusson, L. Guglielmini, S. Lecuyer, R. Rusconi, and H. A. Stone, ``The shape of an elastic filament in a two-dimensional corner flow,'' Phys. Fluids. {\bf 23}, 063602 (2011). 
\bibitem{Wexler2013} J. S. Wexler, P. H. Trinh, H. Berthet, N. Quennouz,
O. du Roure, H. E. Huppert, A. Lindner and H. A. Stone, ``Bending of elastic fibres in viscous flows:
 the influence of confinement,'' J. Fluid Mech.{\bf 720}, 517-544 (2013).
\bibitem{Habibi2007}M. Habibi, N.M. Ribe and D. Bonn, ``Coiling of elastic ropes,'' Phys. Rev. Lett. {\bf99}, 154302  (2007).
\bibitem{Semin2011}B. Semin, H. Auradou and M. Fran\c ois,
``Accurate measurement of curvilinear shapes by Virtual Image Correlation,'' Eur. Phys. J. App. Phys. \textbf{56}, 10701 (2011). 
\bibitem{Berthet2013} H. Berthet, M. Fermigier and A. Lindner, "Single fiber transport in a confined
  channel: microfluidic experiments and numerical study,'' Phys. Fluids \textbf{25}, 103601 (2013).
\bibitem{Richou2004}A. Ben Richou, A. Ambari, and J.K. Naciri, "Drag
force on a circular cylinder midway between two parallel plates at
very low Reynolds numbers. Part 1: Poiseuille flow (numerical),''  Chem.
Eng. Sci. \textbf{59}, 3215-3222 (2004). 
\bibitem{Richou2005} A. Ben Richou, A. Ambari, M. Lebey and J. K. Naciri, ``Drag force on a cylinder midway between two parallel plates at $Re \ll 1$. Part2: moving uniformly (numerical and experimental),'' Chem. Eng. Sci, {\bf{60}}, 2535-2543 (2005).
\bibitem{Semin2009} B. Semin,  J-P. Hulin, and H. Auradou, ``Influence of flow confinement on the drag force on a static cylinder,'' Phys. Fluids \textbf{21}, 103604 (2009).
\end{thebibliography}
\end{document}